\documentclass[12pt,a4paper,final]{iopart}
\usepackage{iopams}  
\usepackage{graphicx}
\usepackage[breaklinks=true,colorlinks=true,linkcolor=blue,urlcolor=blue,citecolor=blue]{hyperref}

\expandafter\let\csname equation*\endcsname\relax
\expandafter\let\csname endequation*\endcsname\relax
\usepackage{amsmath}
\usepackage{graphicx}
\usepackage{amsfonts}
\usepackage{mathtools}
\usepackage{float}
\usepackage{epsfig}
\usepackage{color}
\usepackage{gensymb}
\usepackage{multirow}
\usepackage[mathscr]{euscript}
\usepackage{pifont}

\newcommand{\eee}[1]{\ensuremath{\times 10^{#1}}}
\newcommand{\ee}[1]{\ensuremath{10^{#1}}}
\newcommand{\localdensity}{\eta}
\newcommand{\onlinecite}[1]{\cite{#1}}

\begin{document}

\title{The structure factor of primes}

\author{G. Zhang}

\address{Department of Chemistry, Princeton University,
Princeton NJ 08544 }

\author{F. Martelli}

\address{Department of Chemistry, Princeton University,
Princeton NJ 08544}

\author{S. Torquato}

\address{Department of Chemistry, Department of Physics,
Princeton Institute for the Science and Technology of
Materials, and Program in Applied and Computational Mathematics, Princeton University,
Princeton NJ 08544}

\mailto{torquato@electron.princeton.edu}

\begin{abstract}
Although the prime numbers are deterministic, they can be 
viewed, by some measures, as pseudo-random numbers. In this article, we  numerically study the pair statistics of the primes using statistical-mechanical
methods, especially the structure
factor $S(k)$ in an interval $M \leq p \leq M + L$ with $M$ large, and $L/M$ smaller than unity.  
We show that
the structure factor of the prime-number configurations in such intervals
exhibits well-defined Bragg-like peaks along with a small ``diffuse"
contribution. This indicates that the primes are appreciably more correlated 
and ordered than previously thought. 
Our numerical results definitively suggest an explicit formula for the locations and heights
of the peaks.
This formula predicts infinitely many peaks in any non-zero interval, similar to the behavior of quasicrystals. However, primes differ from quasicrystals in that the ratio between the location of any two predicted peaks is rational. We also show numerically that the diffuse part decays slowly as $M$ and $L$ increases. This suggests that the diffuse part vanishes in an appropriate infinite-system-size limit.

\end{abstract}

\vspace{2pc}

\section{Introduction}
\label{Introduction}

The properties of the prime numbers have been a source of fascination since ancient times. 
Euclid proved that there are infinitely many primes.  Given the first $n$ prime numbers
$p_1, p_2, \cdots, p_n$, the subsequent prime can be found deterministically by sieving \cite{samuel1772sieve}.
Nonetheless, there is no known deterministic formula that
can quickly (polynomial in the number of digits in a prime) generate large numbers that are guaranteed to be prime. 
(So far, the largest known prime is $2^{74,207,281} -1$, which is about 22 million digits long \cite{Mersenne}.)
Let $\pi(x)$ denote the {\it prime counting function}, which gives the number of primes
less than integer $x$. According to the  prime number theorem \cite{hadamard1896distribution}, the prime counting
function in the large-$x$ asymptotic limit is given by
\begin{equation}
\pi(x) \sim \frac{x}{\ln(x)} \qquad (x \rightarrow \infty).
\end{equation}
This means that for sufficiently large $x$, the probability that a randomly
selected integer not greater than $x$ is prime is very close to $1 / \ln(x)$,
which can be viewed as position-dependent number density $\rho(x)$ (number of primes up to $x$ divided
by the interval $x$). This implies that  the primes become sparser as $x$ increases
and hence can be regarded as a statistically inhomogeneous set of points that are located
on a subset of the odd integers.

\begin{figure}[h]
\centerline{\includegraphics[width=0.75\textwidth,clip=]{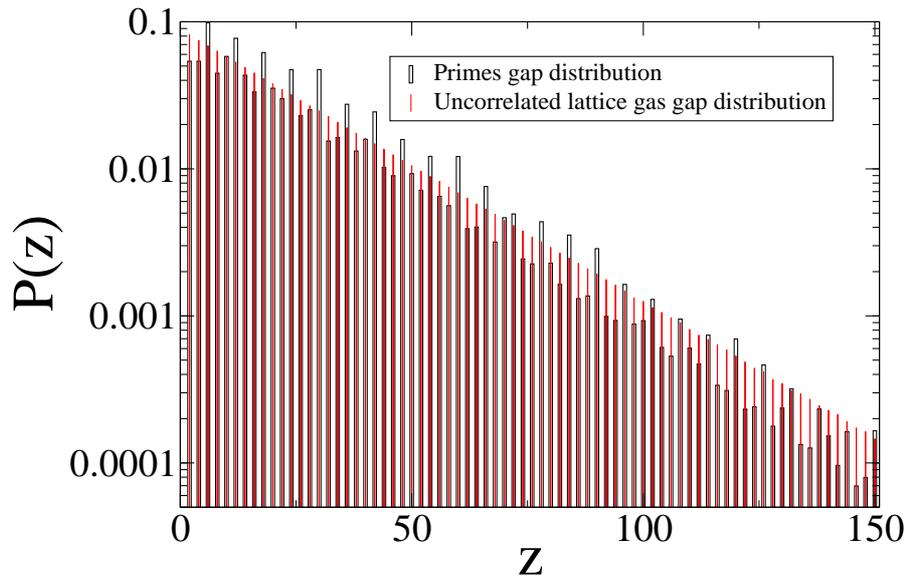}}
\caption{Comparison of the gap distribution for the primes and the uncorrelated
lattice gas with the {same cardinality} (occupation number)  as the set of primes.
The primes are taken to lie on an integer lattice with a spacing of 2, i.e., a
subset of the odd positive integers. We consider $N$ primes in interval $[M,M+L]$ ($M$ large and $M \gg L$).
Here $N=\ee{7}$, $L=244651478$ with $M=42151671493$, the 1,800,000,000th prime number.}
\label{gap}
\end{figure}

While the prime numbers (except for 2) are a deterministic subset of the odd 
integers, they can be viewed, by some measures, as pseudo-random numbers.
Moreover, there are quick {\it stochastic}
ways to find large primes \cite{miller_1976,rabin_1980, pomerance1980pseudoprimes, baillie1980lucas, atkin1993elliptic}, examples of which are based on variants of Fermat's little
theorem~\cite{miller_1976,rabin_1980, pomerance1980pseudoprimes, baillie1980lucas}.
To get a sense of how the primes can be viewed as pseudo-random numbers, let us consider the  {\it gap distribution function} $P(z)$, which gives the probability distribution of the gap size between two consecutive primes, $z$.
Figure \ref{gap} compares the {\it gap probability distribution} $P(z)$ for the primes
to the {\it uncorrelated lattice gas} at the same number density. An ``uncorrelated lattice gas'' refers to a lattice-gas system where each site has a certain probability of being occupied, independent of the occupation of other sites. For an uncorrelated lattice gas at number density $\rho=N/L$ with lattice spacing 2
in the infinite-system-size limit, the gap distribution is exactly given by  
\begin{equation}
P(z)=f(1-f)^{z/2-1},
\end{equation}
where $f=2\rho$ is the probability that a site is occupied.
We see  that by the gap distribution, the primes cannot be clearly distinguished from an uncorrelated lattice gas.
Indeed,  probabilistic methods to treat the primes have yielded fruitful insights
about them \cite{granville1995harald, gallagher1976distribution, montgomery1973pair}. For example, based on the assumption that the primes behave like a Poisson process
(uncorrelated lattice gas), Cram{\'e}r (1920) conjectured that \cite{granville1995harald} for large $x$ 
\begin{equation}
g(x) \ge c \ln^2(x)
\end{equation}
where $g(x)$ denotes the {\it largest prime gap} within an interval $[x,2x]$.

On the other hand, it is known that primes contain unusual patterns.
Chebyshev observed in $1853$ that primes congruent to $3$ modulo $4$ seem to predominate 
over those congruent to $1$ \cite{granville2006prime}. 
Assuming a generalized Riemann hypothesis, Rubinstein and Sarnak ~\cite{rubinstein_1994}
exactly characterized this phenomenon and more general related results. A computational study on the 
Goldbach conjecture demonstrates a  connection based on a modulo $3$ geometry between the set of even integers 
and the set of primes~\cite{martelli_2013}. 
In $1934$, Vinogradov proved that every sufficiently large odd integer is the sum of three primes~\cite{vinogradov_1937}. 
This method has been extended to cover many other types of patterns~\cite{green_2008,green_2006,green_2008_2,tao_2011}. 
Recently it has been shown that there are infinitely many pairs of primes with some finite gap~\cite{zhang_2014}
and that primes ending in $1$ are less likely to be followed by another prime ending in $1$~\cite{lemke_2016}.
Numerical evidence of regularities in the distribution of gaps between primes when these are divided into 
congruence families have also been reported~\cite{maynard_2015,dahmen_2001,wolf_1996}, along with the observation 
of period-three oscillations in the distribution of increments of the distances between consecutive primes
numbers~\cite{kumar_2003}.

The present paper is motivated by certain unusual properties of the 
Riemann zeta function $\zeta(s)$, which is a function of a complex variable $s$
that is intimately related to the primes. The zeta function has many different representations,
one of which is the series formula
\begin{equation}
\zeta(s)=\sum_{n=1}^{\infty} \frac{1}{n^s},
\end{equation}
which only converges for $Re(s)>1$.  However, $\zeta(s)$ has a unique analytic continuation to the entire complex plane,
excluding the simple pole at $s=1$.
According to the {\it Riemann hypothesis}, the nontrivial
zeros of the zeta function lie along the {\it critical line} $s=1/2 + it$ with $t\in\mathbb{R}$ in the complex plane.
The nontrivial zeros tend to get denser the higher on the critical line. When the spacings of the zeros are appropriately
normalized so that they can be treated as a homogeneous point process at unity density, the resulting
pair correlation function takes on the simple form $1-\sin^2(\pi r)/(\pi r)^2$ \cite{montgomery1973pair}. The corresponding
structure factor $S(k)$ (essentially the Fourier transform of $g_2(r)$)  tends to zero 
linearly in the wavenumber $k$ as $k$ tends to zero but is unity for sufficiently
large $k$. This implies that the normalized Riemann zeros possess a remarkable type of correlated
disorder at large length scales known as hyperuniformity~\cite{torquato2008point}. A hyperuniform many-particle system is one in which
the structure factor approaches zero in the infinite-wavelength
limit ~\cite{torquato2003local}. In such systems, density fluctuations are
anomalously suppressed at very large length scales, a ``hidden"
order that imposes strong global structural constraints.
All structurally perfect crystals and quasicrystals are hyperuniform,
but typical disordered many-particle systems, including gases, liquids, and glasses, are not. Disordered
hyperuniform many-particle systems are exotic states of
amorphous matter that have attracted considerable recent attention \cite{  donev2005unexpected, florescu2009designer, zachary2009hyperuniformity, zachary2011hyperuniform, kurita2011incompressibility, xie2013hyperuniformity, jiao2014avian, lesanovsky2014out, muller2014silicon, hexner2015hyperuniformity, jack2015hyperuniformity, de2015toward, torquato2015ensemble, We15, torquato2016hyperuniformity, Le16,xu2016influence, ma2017random}.
The zeta function is directly related to the primes via the following Euler product formula:
\begin{equation}
\zeta(s) =\Big[\prod_{n=1}^{\infty} [1-1/p_n^s]\Big]^{-1},
\end{equation}
which leads to a variety of {\it explicit} formulas that link the primes
on the one hand to the zeros of the zeta
function on the other hand \cite{Da80,Te95,Iw04}.
Thus, one can in principle deduce information about  primes from information about  zeros of the zeta function.
Accordingly, one might expect the primes to
encode hyperuniform correlations that are seen in the Riemann zeros.

In this article, we  numerically study the pair statistics of configurations of the primes, especially the structure
factor $S(k)$ in an interval  $M \leq p \leq M + L$ with $M$ large, and $L/M$ smaller than unity.
As we will detail in Sec.~\ref{sec:Procedure}, this choice of intervals allow us to obtain prime configurations with virtually
constant density from one end of the interval to the other.
We show in Sec. \ref{sec_Numerical} that
the structure factor exhibits well-defined Bragg-like peaks along with a small ``diffuse"
contribution. This indicates that the primes are appreciably more correlated than anyone has previously conceived.
Our numerical results definitively suggest an explicit formula for the locations and heights
of the peaks.
This formula predicts infinitely many peaks in any non-zero interval. Although such a behavior is similar to that of quasicrystals \cite{levine1984quasicrystals, socolar1986quasicrystals, ouguz2017hyperuniformity}, primes differ from the latter in that the ratio between the locations of any two predicted peaks is rational. We also show numerically that the diffuse part decays slowly as $M$ and $L$ increases. In a subsequent paper \cite{To17}, we prove this to be true and investigate its consequences. This might indicate that the diffuse part vanishes
in an appropriate infinite-system-size
limit. 
We note that the structure factor has been used to study other number-theoretic
systems \cite{miller_1976, ba00, torquato2008point, ba11}.  

The rest of the paper is organized as follows:
In Sec. 2, we present relevant definitions and describe the
simulation procedure. In Sec. 3,
we present results for the pair statistics of the primes in certain
intervals in both direct and reciprocal spaces. In Sec. 4, we make concluding remarks.

\section{Definitions and Simulation Procedure}
\label{preliminaries}

\begin{figure}[h]
\includegraphics[width=0.9\textwidth]{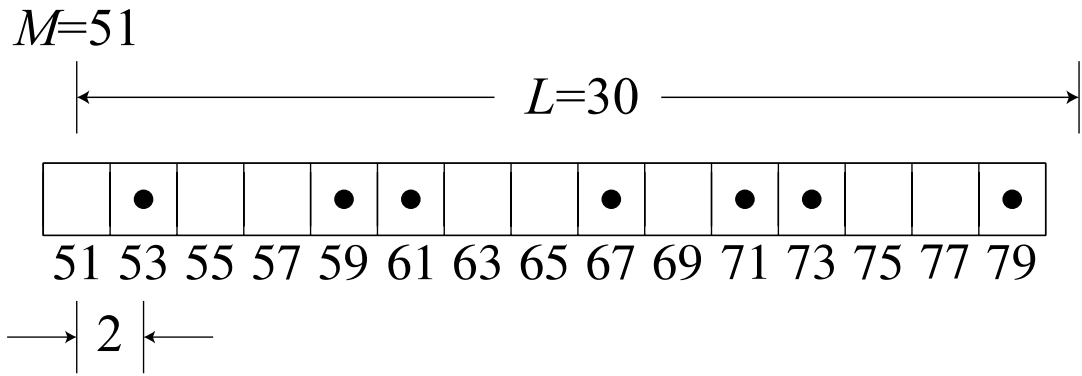}
\caption{Schematic plot  of a prime-number configuration with $M=51$ and $L=30$. Since we always use $M\ge 3$, any prime number in the interval $[M, M+L)$ is odd. Therefore, a prime-number configuration is a lattice gas with lattice spacing 2 in which the primes are the ``occupied" sites
and the composites are ``unoccupied" sites.}
\label{cartoon}
\end{figure}

We will study the pair correlation function as well as the structure factor of prime-number configurations.  
Similar to Ref.~\onlinecite{Ve13}, we treat the primes in some interval $[M,M+L]$ to be a special lattice-gas model: the primes are  
``occupied'' sites on a integer lattice of spacing 2 that contains all  of the positive odd integers and 
the unoccupied sites are the odd composite integers.
As detailed below, we consider the positive parameter $L/M$ generally to be smaller than unity to ensure homogeneity.
Within this interval, let $N_s$ and $N$ be the total number of sites and the total number of primes, respectively.
In practice, we consider the half-open interval $[M,M+L)$ because it provides a simple relation
between $L$ and $N_s$, namely, $L=2N_s$.
 For all cases considered in this paper, $M\ge 3$.
Figure~\ref{cartoon} illustrates an example of a prime configuration. 
The rest of this section details the mathematical tools that we use to treat such systems.
 
\subsection{Discrete Fourier Transform}
For a function $f(r)$ defined on an integer lattice with spacing $a$ 
that is contained within a periodic box of length  $L$, one may define its Fourier transform
as follows:
\begin{equation}
{\tilde f}(k)=\sum_{r=0, a,2a,\cdots,L-a} f(r) \exp(ikr),
\label{fft}
\end{equation}
where the parameter $k$ is an integer multiple of $2\pi/L$. 
The inverse transform is given by
\begin{equation}
f(r)=\frac{1}{N_s} \sum_{j=0}^{N_s-1} {\tilde f}\left(\frac{2\pi j}{L}\right)\exp\left(-i \frac{2\pi j}{L} r\right),
\label{ift}
\end{equation}
where $N_s=L/a$ is the number of sites.

\subsection{Pair Statistics and their basis properties}

We define $\localdensity(r)$ as the indicator function such that $\localdensity(r)=1$ if the site at $r$ is occupied, and $\localdensity(r)=0$ otherwise.
Let ${\tilde \localdensity}(r)$ be its Fourier transform. 
We define occupation fraction to be $f=<\localdensity(r)>$, where $<>$ denotes an average over all $r$.
We define the structure factor as
\begin{equation}
S(k) = |{\tilde \localdensity}(k)|^2/N-N \delta_{k, 0}.
\label{sk}
\end{equation}
Define the pair correlation function $g_2(r)$ as 
\begin{equation}
g_2(r)=\frac{1}{Nf}\sum_{n=0, a,2a,\cdots,L-a}\localdensity(n)\localdensity(n+r) - \frac{\delta_{r, 0}}{f}.
\label{PairCorrelation}
\end{equation}
By definition, $g_2(0)=0$. For $r \neq 0$, $g_2(r)$ can be interpreted as the probability that the site at $p+r$ is occupied given that the site at $p$ is occupied divided by $f$.

The structure factor and the pair correlation function are related as follows:
\begin{equation}
\begin{split}
N \delta_{k, 0}+S(k) &= |{\tilde \localdensity}(k)|^2/N\\
&=\frac{1}{N}\sum_{m=0, a,2a,\cdots,L-a}\localdensity(m)\exp(ikm)\sum_{n=0, a,2a,\cdots,L-a}\localdensity(n)\exp(-ikn)\\
&= \frac{1}{N}\sum_{m=0, a,2a,\cdots,L-a} \hspace{100\in}\sum_{n=0, a,2a,\cdots,L-a}\localdensity(m)\localdensity(n)\exp[ik(m-n)]\\
&=\frac{1}{N}\sum_{r=0, a,2a,\cdots,L-a} \hspace{100\in}\sum_{n=0, a,2a,\cdots,L-a}\localdensity(n+r)\localdensity(n)\exp(ikr)\\
&= \frac{1}{N}\sum_{n=0, a,2a,\cdots,L-a}\localdensity(n+0)\localdensity(n)\exp(ik0) \\&+ \frac{1}{N}\sum_{r=a,2a,\cdots,L-a} \hspace{100\in}\sum_{n=0, a,2a,\cdots,L-a}\localdensity(n+r)\localdensity(n)\exp(ikr)\\
&=\frac{N}{N}+\frac{1}{N}\sum_{r=a,2a,\cdots,L-a} Nfg_2(r)\exp(ikr)\\
&=1+f\sum_{r=a,2a,\cdots,L-a} g_2(r)\exp(ikr)\\
&=1+f\sum_{r=0, a,\cdots,L-a} g_2(r)\exp(ikr)\\
\end{split}
\label{ft}
\end{equation}
This equation enables us to obtain a sum rule for both $g_2(r)$ and $S(k)$. For $g_2(r)$, plugging $k=0$ into Eq.~(\ref{ft}) yields
\begin{equation}
\frac{N-1}{f}=\sum_{r=a,2a,\cdots,L-a} g_2(r).
\end{equation}
The sum rule for $S(k)$ is easily found by invoking the inverse Fourier transform equation:
\begin{equation}
g_2(r)=\frac{1}{N_sf} \sum_{j=0}^{N_s-1} \left[S\left(\frac{2\pi j}{L}\right)+N \delta_{k, 0}-1\right]\exp\left(-i \frac{2\pi j}{L} r\right).
\end{equation}
At $r=0$, this relation becomes
\begin{equation}
\begin{split}
0&= \sum_{j=0}^{N_s-1} \left[S\left(\frac{2\pi j}{L}\right)+N \delta_{k, 0}-1\right]\\
&=\sum_{j=1}^{N_s-1} S\left(\frac{2\pi j}{L}\right)+N - N_s,
\end{split}
\end{equation}
and hence the sum rule for the structure factor is given by
\begin{equation}
\sum_{j=1}^{N_s-1} S\left(\frac{2\pi j}{L}\right)=N_s-N.
\label{SkSum}
\end{equation}
For the primes, $L=2N_s$, and hence the sum rule is specifically
\begin{equation}
\sum_{j=1}^{N_s-1} S\left(\frac{\pi j}{N_s}\right)=N_s-N.
\end{equation}

The fact that all primes greater than 3 are odd integers lead to a few important properties of $S(k)$. First, $N \delta_{k, 0}+S(k)$ is a periodic function of period $\pi$ ,since
\begin{equation}
\begin{split}
N \delta_{k+\pi, 0}+S(k+\pi)&=\frac{|\sum_{j=1}^N \exp[-i (k+\pi) (p_j-M)]|^2}{N}\\ &=\frac{|\sum_{j=1}^N \exp[-i k (p_j-M)]\exp(-i \pi (p_j-M)]|^2}{N} \\ &=\frac{|-\exp(i \pi M)\sum_{j=1}^N \exp[-i k (p_j-M)]|^2}{N} \\ &=\frac{|\sum_{j=1}^N \exp[-i k (p_j-M)]|^2}{N} \\ &=N \delta_{k, 0}+S(k),
\end{split}
\label{period}
\end{equation}
Second, from Eqs.~(\ref{fft})~and~(\ref{sk}), one can see that when $k=m\pi$, where $m$ is any non-zero integer, $S(k)=N$ achieves the global maximum of this function. 
The function $S(k)$ therefore displays strong peaks at such $k$ values.
Third, from Eqs.~(\ref{fft})~and~(\ref{sk}), one can see the function $S(k)$ has reflection symmetry $S(k)=S(-k)$. This reflection symmetry, combined with the periodicity [Eq.~(\ref{period})], implies another reflection symmetry, $S(\pi/2+k)=S(\pi/2-k)$. With these properties in mind, we only need to study $S(k)$ in the range $0<k\le \pi$ in this paper.

\subsection{Simulation Procedure}
\label{sec:Procedure}


In statistical mechanics, the study of $S(k)$ often focuses on statistically homogeneous systems. However, the prime numbers are not homogeneous. Instead, in the vicinity  of $x$, the density of prime numbers scales as $1/\ln(x)$. To overcome this difference, we focus on large $M$ values and let $L/M$ be a constant less than unity
\footnote{Strictly speaking, the requirement that $L/M<1$ is not important. Densities from one end to the other would be nearly constant in the $M \to \infty$ limit even if $L/M=\beta>1$. However, for computationally practical values of $M$, a much smaller $L$ is desired to accurately ascertain constant density.}. 
This implies that the ``local" density from the beginning to the end of the  interval $[M, M+L)$   is
virtually a constant. For example, we will study a system of $M=\ee{10}$ and $L=\ee{7}$. As $x$ changes from $M$ to $M+L$, $1/\ln(x)$ only changes from 0.043429$\ldots$ to 0.043427$\ldots$. 

\begin{figure}[h]
\includegraphics[width=0.7\textwidth]{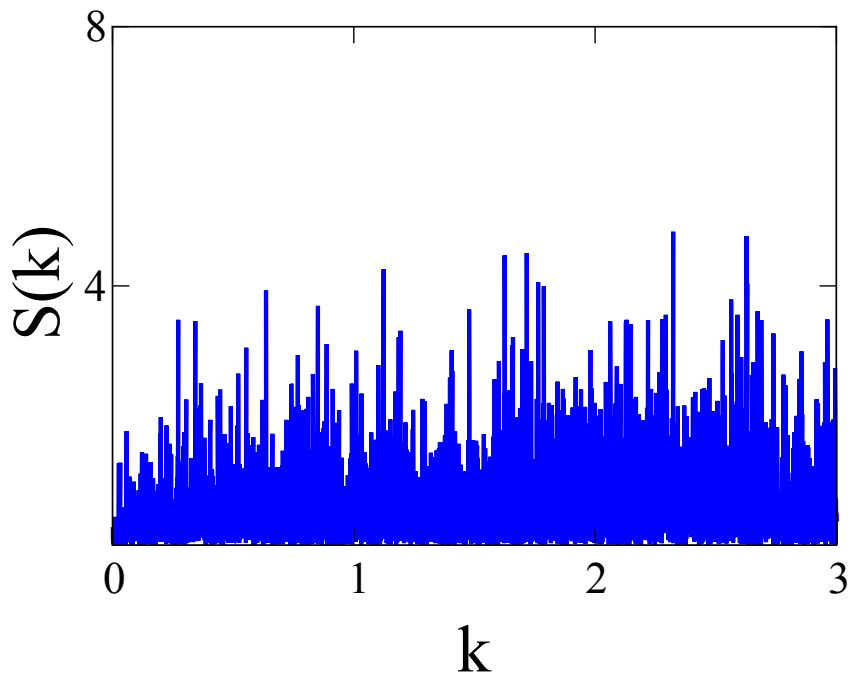}
\caption{Structure factor $S(k)$ associated with $p/\ln(p)$ for all prime number $p$'s in the interval $[3, 3+\ee{5})$.}
\label{RescaledSk}
\end{figure}

We minimize the problem of inhomogeneity by requiring sufficiently large $M$. Instead, one might think an even better solution to this problem is to rescale the configuration such that it is homogeneous.
The natural scaling is to replace each prime number $p$ with $p/\ln(p)$. However, it turns out that after performing such rescaling, the structure 
factor appears to be 
completely noisy with no obvious peaks with heights
comparable to N; see Fig.~\ref{RescaledSk}
\footnote{Notice that in this interval, primes possess a sharp density gradient, and the $p \to p/\ln(p)$ rescaling is strongly non-linear. If we had chosen an interval with a negligible density gradient, then the $p \to p/\ln(p)$ rescaling would be almost linear, and $S(k)$ would  essentially be a simple rescaling of the ones reported below (Fig.~\ref{sk0}).}. 
This is to be contrasted
with our findings reported in the rest of the paper, in which we choose to study the primes in the interval 
$[M,M+L)$ with $M$ large, and $L/M$ smaller than unity.

We obtain a list of prime numbers from Ref.~\onlinecite{PrimesList},
and calculate $S(k)$ of prime numbers and uncorrelated lattice gases with the fast Fourier transform (FFT) algorithm using the kissFFT software \cite{kissFFT}. This algorithm has the advantage of not only being ``fast'' \footnote{The time complexity of calculating $S(k)$ for all $k$'s using FFT algorithm scales as $L\log(L)$, while the time complexity of doing so using Eq.~(\ref{ft}) scales as $LN$.}, but also being accurate, as the upper bound on the relative error scales as $\epsilon \log(L)$, where $\epsilon$ is the machine floating-point relative precision. We use double-precision numbers to further minimize $\epsilon$.

More precisely, FFT allows one to efficiently calculate
\begin{equation}
X(q)=\sum_{n=0}^{T-1} x_n \exp(-2\pi i qn/T)
\end{equation}
for arbitrary $x_0$, $x_1$, $\cdots$, $x_{T-1}$. Here, we simply let $T=L/2$, and let $x_j=1$ if $M+2j$ is a prime number and $x_j=0$ if $M+2j$ is composite. The structure factor is then calculated from:
\begin{equation}
S(2\pi q/L)=|X(q)|^2/N.
\end{equation}

\section{Results for the Pair Statistics of the Prime Numbers}
\label{sec_Numerical}


\begin{figure}
\includegraphics[width=0.6\textwidth]{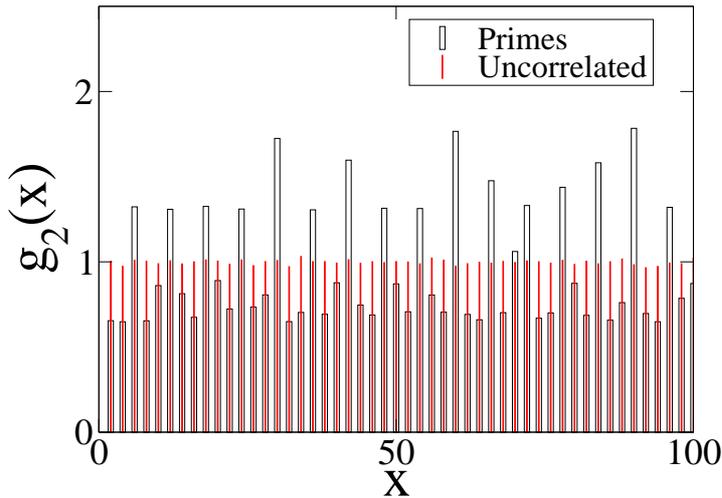}
\caption{Pair correlation function $g_2(x)$, as defined in Eq.~(\ref{PairCorrelation}), for a prime number configuration of $M=\ee{10}+1$, $L=\ee{6}$, and $N=43427$, compared with $g_2(x)$ of a uncorrelated lattice gas configuration of the same $L$ and $N$.}
\label{fig_G2}
\end{figure}

The pair correlation function as defined in Eq.~(\ref{PairCorrelation}), $g_2(x)$, of prime numbers is presented in Fig.~\ref{fig_G2} and compared with $g_2(x)$ of uncorrelated lattice gases. This quantity for uncorrelated lattice gas is simply
\begin{equation}
g_2(x)=\frac{N-1}{f(N_s-1)}
\end{equation}
for any $x\neq 0$. This is because after one site is occupied, out of the remaining $N_s-1$ sites, exactly $N-1$ sites are occupied. We see that by this measure, the prime numbers appear to be distinctly different from uncorrelated lattice gases. We see that $g_2(x)$ for primes is higher than $g_2(x)$ of the uncorrelated lattice gas if and only if $x$ is divisible by 3. However, we will see that the difference in pair statistics is much more obvious when we study $S(k)$ below.

\begin{figure}[h]
\includegraphics[width=0.5\textwidth,clip=]{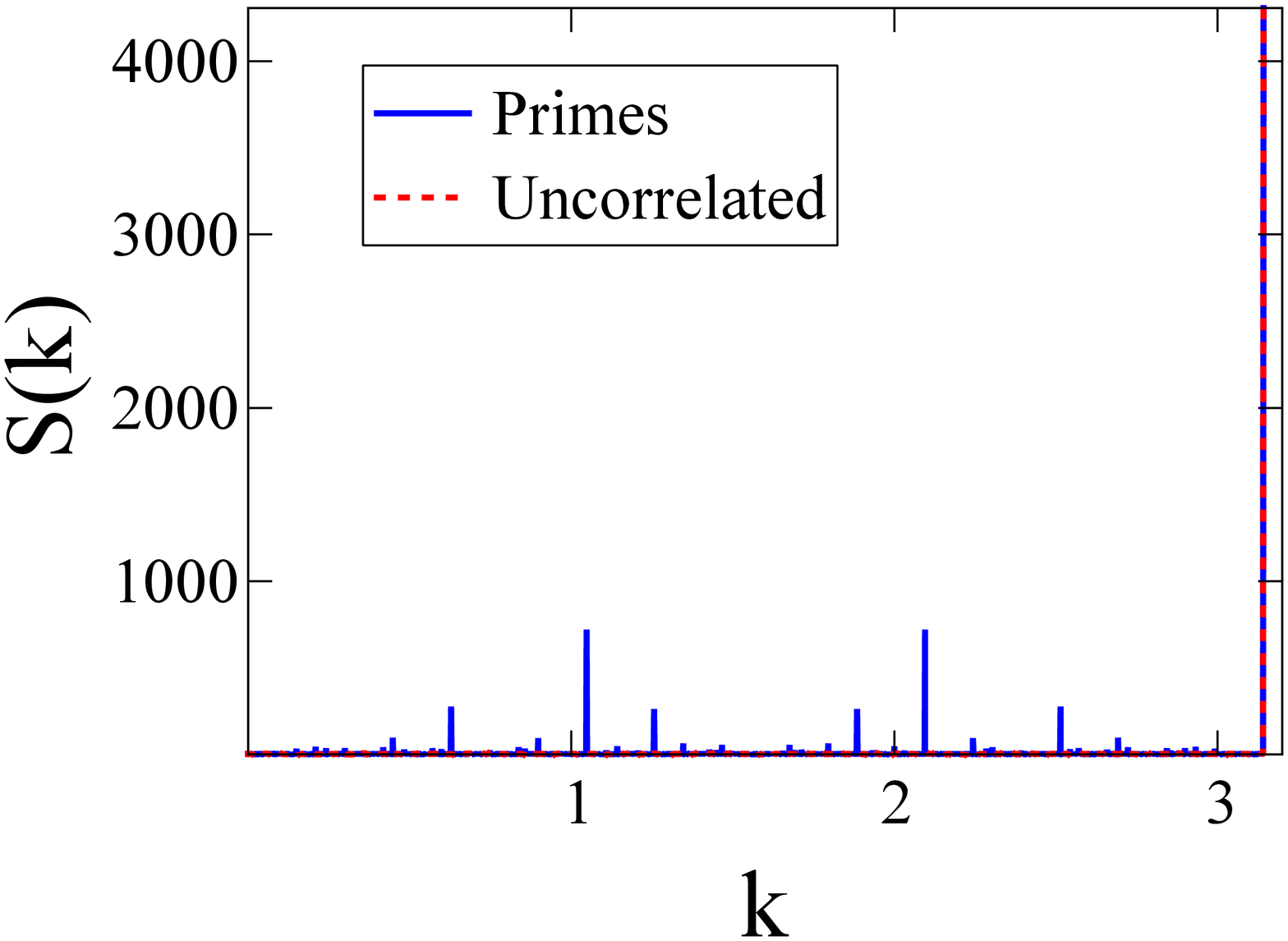}\hspace{0.25in}
\includegraphics[width=0.4\textwidth,clip=]{Fig5b.eps}
\caption{Left: $S(k)$ for prime numbers as a function of $k$ (in units of the integer lattice spacing) for $M=\ee{10}+1$ and $L=\ee{5}$ contains many well-defined Bragg-like peaks of various heights, creating a type of self-similarity. Right: A zoomed-in view revealing the existence of a small, noisy ``diffuse part'' besides the peaks. We also plot $S(k)$ for uncorrelated lattice gases for comparison. As we have discussed in Sec.~\ref{preliminaries}, we only show $S(k)$ in the range $0<k \le \pi$, and therefore omit the peak at $k=0$.
}
\label{sk0}
\end{figure}

We present and study numerically calculated structure factors of prime numbers for various $M$'s and $L$'s in this section. At first, let us examine $S(k)$ for $M=\ee{6}+1$ and $L=5000$, which is presented in Fig.~\ref{sk0}. At a larger scale, $S(k)$ appears to consist of many well-defined Bragg-like peaks of various heights, with the highest peak occurring at $k=\pi$ (left panel). As we zoom in, it becomes evident that besides those peaks, $S(k)$ also has a random, noisy contribution that is often below 1 (right panel). We will call the latter contribution the ``diffuse part'' in the rest of the paper. Figure~\ref{sk0} also includes $S(k)$ for uncorrelated lattice gases,
which consists of a diffuse part and a single peak at the trivial value of $k=\pi$. Away from the peak, $S(k)$ for uncorrelated lattice gases fluctuates around an average value of $\frac{N_s-N}{N_s-1}$ [this particular average value is required by the sum rule, Eq. (\ref{SkSum})]. A major conclusion is that the structure factor
of the primes is characterized by a substantial amount of order across length scales, relative to the uncorrelated lattice
gas, as evidenced by the appearance of many Bragg-like peaks.
At this stage, it seems that the existence of the diffuse part makes primes non-hyperuniform. However, we will show in Sec.~\ref{sec:diffuse} that the diffuse part decreases as $L$ increases, and suggest that it vanishes in the infinite-system-size limit.

At this stage, the distinction between the peaks and the ``diffuse part'' is somewhat unclear. Since $S(k)$ contains peaks of various heights, is it possible that the diffuse part is actually made of many smaller peaks? We can only answer this question after we study the peaks and the diffuse part more deeply later in this section.

\subsection{Peaks}
\label{sec:Peaks}

We move on to study the peaks. From Fig.~\ref{sk0}, one sees that the highest peak is at $k=\pi$, which is trivially caused by the periodicity of the underlying lattice. The next highest two peaks are at $k=\pi/3$ and $k=2\pi/3$ \footnote{It should be noted that since $k$ has to be integer multiples of $2\pi/L$, for $L=\ee{5}$, $k=\pi/3$ and $k=2\pi/3$ cannot be chosen. The actually observed peaks occur at the closest allowed $k$ points instead.}. Even lower peaks occur at $k=\pi/5$, $2\pi/5$,  $3\pi/5$, and $4\pi/5$. Still lower peaks occur 
at $k=\pi/7$, $2\pi/7$,  $3\pi/7$, $4\pi/7$, $k=5\pi/7$, and $6\pi/7$. Examining $S(k)$ of a much larger system ($M=\ee{10}+1$ and $L=\ee{7}$) 
revealed that there are even more peaks with locations that obey the formula $k=m \pi/n$, where $m$ is any integer coprime with $n$ and $n$ is any square-free odd integer and hence has a distinct prime factorization, i.e., $n=\displaystyle \prod_{j=1}^J p_j$, where $J$ is a positive integer, and $p_1$, $p_2$, $\cdots$, $p_J$ are non-repeating prime numbers larger than 2. If $n$ is even or is not square-free, then we observe no peak at $k=m \pi/n$. We verified the existence of such peaks for $n$ up to $300$. As $n$ increases beyond 300, however, the peaks become too weak to be distinguishable from the diffuse part.

\begin{figure}[h]
\includegraphics[width=0.7\textwidth]{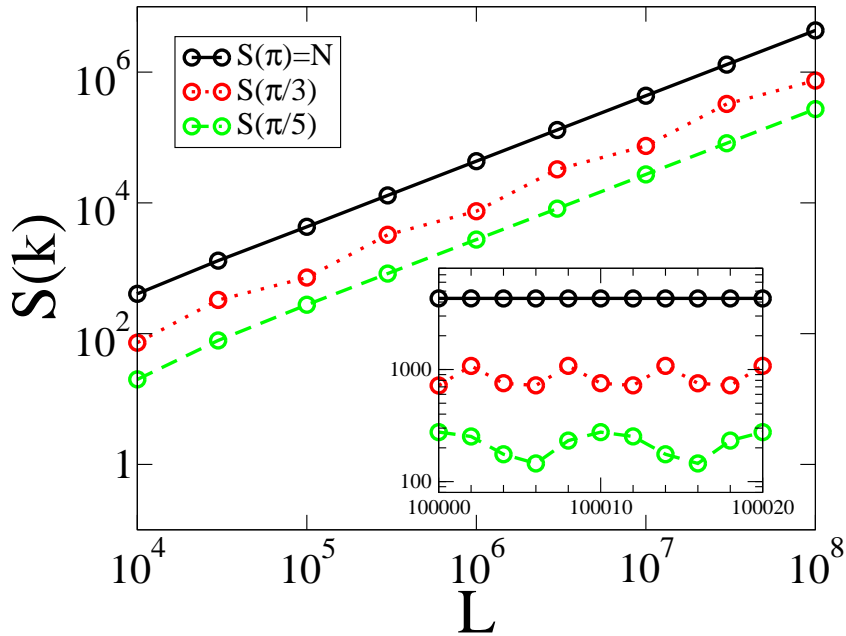}
\caption{The structure factor $S(k)$ at $k=\pi$, $k=\pi/3$, and $k=\pi/5$, as a function of $L$ at $M=\ee{10}+1$. The inset presents more data for $\ee{5}\le L \le \ee{5}+20$.}
\label{peakheights}
\end{figure}

Having an analytical formula of the peak locations, we move on to study the peak heights. As we have shown earlier, the height of the peak at $k=\pi$ is simply $N$. What can we say about the heights of the other peaks? In Fig.~\ref{peakheights} we present computed peak heights at $k=\pi/3$ and $k=\pi/5$ for $M=\ee{10}+1$ and various $L$'s. We see that as $L$ grows, the heights of the peaks at $k=\pi/3$ and $k=\pi/5$ also grow and remain roughly proportional to $N$. Looking at the inset, we see that both $S(\pi/3)$ and $S(\pi/5)$ oscillates periodically as $L$ increases: $S(\pi/3)$ attains a maximum when $L$ is divisible by 3, and $S(\pi/5)$ attains a maximum when $L$ is divisible by 5. Examining the heights of other peaks, we find that the height of a peak at $k=m \pi/n$ is indeed highest when $L$ is divisible by $n$.

Since the divisibility of $L$ with $n$ affects the peak heights and hence will introduce unintended errors if not chosen properly, we desire an $L$ that is divisible by as many prime numbers as possible. We therefore chose $L=2 \times  3 \times  5 \times    7   \times 11   \times 13   \times 17  \times  19=9699690$ and recomputed the heights of several peaks. The results are summarized in Table~\ref{peakheightsTable}. We find that when $L$ is divisible by $n$, the height of the peak at $k=m \pi/n$ is very close to $\displaystyle N\prod_{j=1}^J (p_j-1)^{-2}$, where $p_j$ are the distinct prime factors of $n$.

\begin{table}
\centering
\caption{Peak heights at several different $n$ and $m$'s for $M=2.5\eee{8}+1$ and $L=9699690$ and comparison with the predicted height from the analytical formula.}
\label{peakheightsTable}
\renewcommand{\arraystretch}{0.6}
\begin{tabular}{|l|l|l|l|}
\hline
$n$                  & $m$ & $S(m\pi/n)/N$ & Postulated analytical formula                                 \\ \hline
3                    & 1   & 0.2500000003  & $(3-1)^{-2}=0.25$                                             \\ \hline
\multirow{2}{*}{5}   & 1   & 0.06268293536 & \multirow{2}{*}{$(5-1)^{-2}=0.0625$}                          \\ \cline{2-3}
                     & 2   & 0.06231833526 &                                                               \\ \hline
\multirow{3}{*}{7}   & 1   & 0.02764696627 & \multirow{3}{*}{$(7-1)^{-2}=0.02777\cdots$}                   \\ \cline{2-3}
                     & 2   & 0.02783423055 &                                                               \\ \cline{2-3}
                     & 3   & 0.02785282486 &                                                               \\ \hline
\multirow{4}{*}{15$=3\times5$}  & 1   & 0.01564115190 & \multirow{4}{*}{$[(3-1)(5-1)]^{-2}=0.015625$}                 \\ \cline{2-3}
                     & 2   & 0.01583266309 &                                                               \\ \cline{2-3}
                     & 4   & 0.01551814312 &                                                               \\ \cline{2-3}
                     & 7   & 0.01550964066 &                                                               \\ \hline
\multirow{24}{*}{105$=3\times5\times7$} & 1   &0.0004096963803& \multirow{3}{*}{$[(3-1)(5-1)(7-1)]^{-2}=0.00043402777\cdots$} \\ \cline{2-3}
                     & 2   &0.0004418025682&                                                               \\ \cline{2-3}
                     & 4   &0.0004305924622&                                                               \\ \cline{2-3}
                     & 8   &0.0003879974866&                                                               \\ \cline{2-3}
                     & 11  &0.0004203223484&                                                               \\ \cline{2-3}
                     & 13  &0.0004411107279&                                                               \\ \cline{2-3}
                     & 16  &0.0004191249498&                                                               \\ \cline{2-3}
                     & 17  &0.0003893716268&                                                               \\ \cline{2-3}
                     & 19  &0.0004388128207&                                                               \\ \cline{2-3}
                     & 22  &0.0004193036024&                                                               \\ \cline{2-3}
                     & 23  &0.0004375599695&                                                               \\ \cline{2-3}
                     & 26  &0.0004203613535&                                                               \\ \cline{2-3}
                     & 29  &0.0004418187004&                                                               \\ \cline{2-3}
                     & 31  &0.0004457650582&                                                               \\ \cline{2-3}
                     & 32  &0.0004237635619&                                                               \\ \cline{2-3}
                     & 34  &0.0004500979160&                                                               \\ \cline{2-3}
                     & 37  &0.0004466597486&                                                               \\ \cline{2-3}
                     & 38  &0.0004663304920&                                                               \\ \cline{2-3}
                     & 41  &0.0004255673779&                                                               \\ \cline{2-3}
                     & 43  &0.0004845933410&                                                               \\ \cline{2-3}
                     & 44  &0.0004679589962&                                                               \\ \cline{2-3}
                     & 46  &0.0004572985410&                                                               \\ \cline{2-3}
                     & 47  &0.0004095637658&                                                               \\ \cline{2-3}
                     & 52  &0.0004521962772&                                                               \\ \hline
\end{tabular}
\end{table}

\begin{figure}[h]
\includegraphics[width=0.7\textwidth]{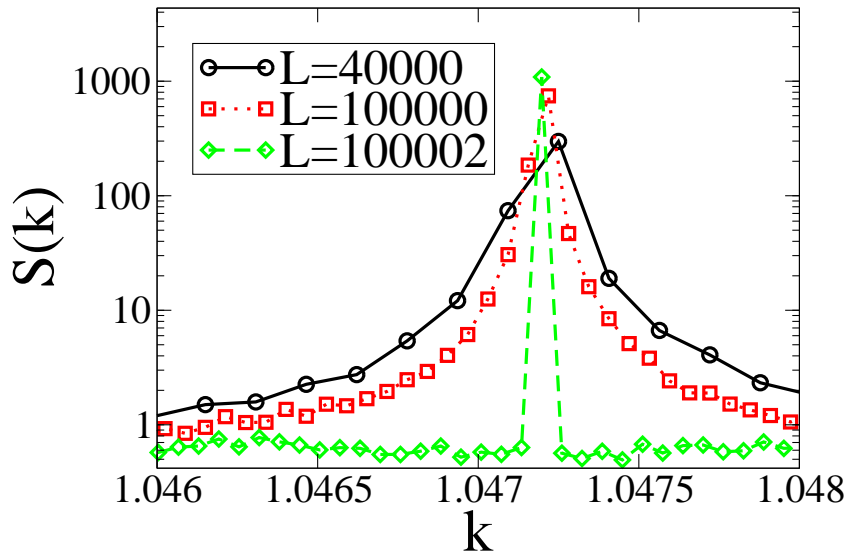}
\caption{$S(k)$ near $k=\pi/3$ for three different $L$'s. Each curve is averaged over 100 prime-number configurations, with the $j$th configuration consists of all prime numbers in the range $[\ee{10}+(j-1)L+1, \ee{10}+jL+1)$.}
\label{peakshape}
\end{figure}

Do these numerically generated peaks have finite or infinitesimal width? To answer this question, we present a close view of the peak at $k=\pi/3$ for three different $L$'s in Fig.~\ref{peakshape}. It turns out that, if $L$ is divisible by $n$, the peak at $k=m\pi/n$ has infinitesimal width, in the sense that $S(k)$ at one $k$ value attains the local maximum and $S(k)$ at all adjacent $k$ values are as low as the typical diffuse part. However, if $L$ is not divisible by $n$, then the peak has a finite width, as $S(k)$ of all $k$ values very close to $m\pi/n$ rises and become much higher than the typical diffuse part. In Fig.~\ref{peakshape} one can also see that when peak widths are finite, a lower $L$ results in a more broadly spread peak. Therefore, all of the peaks may have infinitesimal width in the infinite-$L$ limit. However, in a finite-$L$ simulation, choosing an $L$ that is divisible by as many prime numbers as possible provides a better estimate of $S(k)$ in the infinite-$L$ limit. As $n$ increases, one finds an increasing number of lower peaks, resulting in a statistical self-similarity.

\subsection{Diffuse Part}
\label{sec:diffuse}

\begin{figure}[h]
\includegraphics[width=0.7\textwidth]{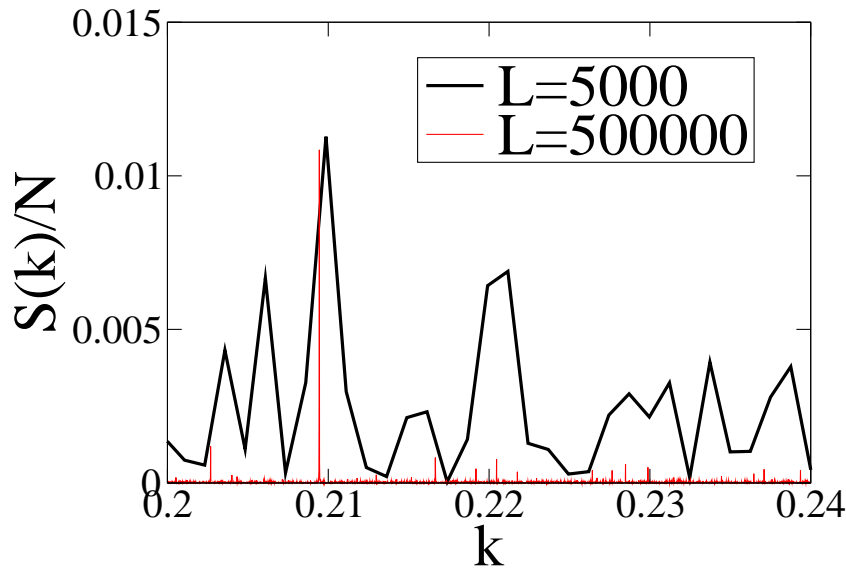}
\caption{The structure factor, $S(k)$, normalized by $N$, for two different $L$'s and $M=\ee{6}+1$.}
\label{differentL}
\end{figure}

The above analysis suggest that the structure factor of the primes possesses infinitely many Dirac-delta-function peaks of various
heights in the infinite-system-size limit. Therefore, one might naturally ask, could the random, noisy ``diffuse part'' be simply a superposition of many small peaks? The answer is no. In Fig.~\ref{differentL} we present $S(k)$ of two different $L$'s in the range $0.2\le k\le 0.24$. For $L=5000$, $S(k)$ in this range appeared completely random and noisy, matching our definition of the diffuse part. To see if this diffuse part is actually a superposition of many small peaks, we compare it to the structure factor for $L=500000$. The larger $L$ allows more $k$ points to be chosen, and therefore improves the $k$ resolution, and reveals some peaks in this $k$ range. We see that although $S(k)$ of the smaller $L$ appeared to be entirely random, the maximum at $k \approx 0.21$ corresponds to a strong peak of the larger system, and is therefore actually a peak. However, other maxima for the smaller system do not correspond to peaks for the larger system, and can only be explained by assuming the existence of a noisy contribution to $S(k)$, which we call the ``diffuse part.'' To summarize, in Fig.~\ref{differentL} we show that for finite systems, there is clearly a noisy contribution to $S(k)$ other than the peak contribution, even though distinguishing these two components of $S(k)$ can be difficult without consulting the analytical formula for the peaks.

\begin{figure}
\includegraphics[width=0.7\textwidth]{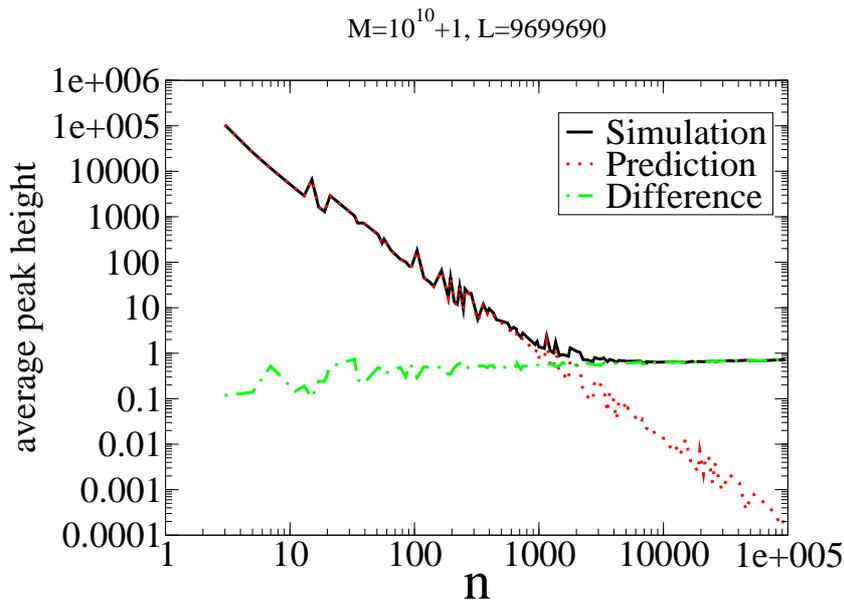}
\caption{Average peak height of all peaks of a given $n$, the predicted peak heights, and their difference for all $n<\ee{5}$ that are odd, square-free, and divide $L$ evenly. Here $M=\ee{10}+1$ and $L=9699690$. For each $n$, we find all $m$'s that are coprime with $n$, and average the heights of peaks at $m\pi/n$. The average turns out to be always greater than the prediction, $\displaystyle N\prod_{j=1}^J (p_j-1)^{-2}$. Their difference is between 0.1 and 1, which is of the same order of magnitude as the diffuse part.}
\label{PeakDifference}
\end{figure}

The diffuse part contributes to not only $k$ points where there are no peaks, but also to $k$ points where there are peaks. In Fig.~\ref{PeakDifference}, we compare numerical peak heights, averaged over all allowed $m$ for a particular $n$, with the analytic peak formula described in Sec.~\ref{sec:Peaks}. It turns out that the numerical average is always slightly higher than that predicted by this formula, and their difference is of the same order of magnitude as the diffuse part. Thus, $S(k)$ at predicted peak locations is actually the sum of the peak and diffuse contributions. 

\begin{figure}[h]
\includegraphics[width=0.7\textwidth]{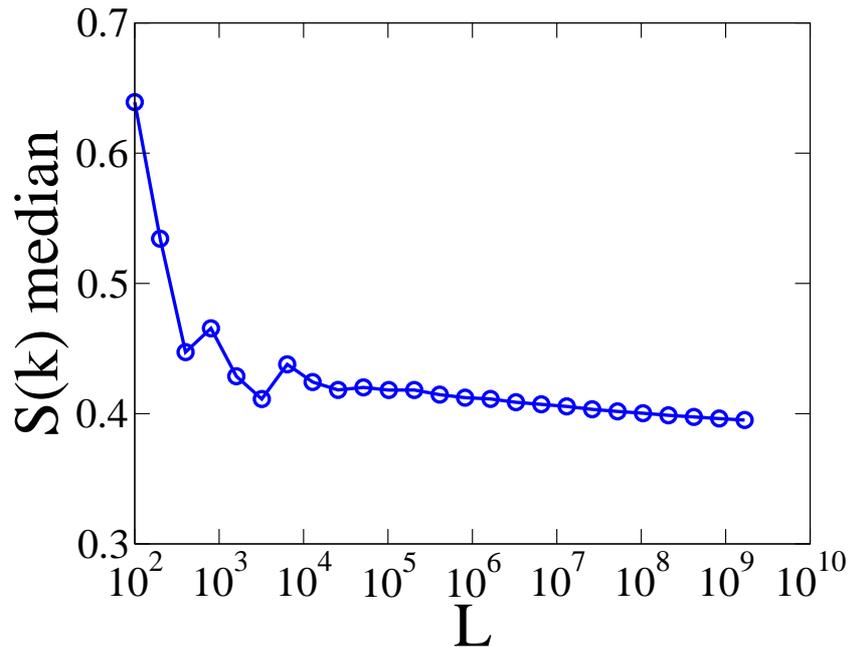}
\caption{The median of the structure factor, $S(k)$, for all possible choices of $k$, as a function of $L$. Here $M$ is chosen to be $10L$. 
The diffuse part of primes appears to be slowly decreasing as $L$ increases. This is to be contrasted with the
 uncorrelated lattice gas with an appreciably larger predictable diffuse part in which there is no dependence on the system size.}
\label{median}
\end{figure}

We can quantify the diffuse part using the median of $S(k)$ for all possible choices of $k$ in the range $0<k\le \pi$. We present these medians in Fig.~\ref{median}. As $L$ increases, the median of $S(k)$ generally decreases. However, with our current data, it is unclear if the median of $S(k)$ approaches zero in the $L \to \infty$ limit.

\section{Conclusions}

In summary, we have numerically shown that the structure factor of prime numbers in the interval 
$M \leq p < M + L$ with large $M$ and $L/M < 1$ exhibit well-defined Bragg-like peaks, alongside a very small diffuse part, which slowly decays as the system size increases. In contrast, the peaks persist as the system size increases. Therefore, we have numerically shown in this paper that the primes are characterized by a substantial amount of order, especially relative to an uncorrelated lattice gas that does not have any such peaks. 
We also show that peaks at $k=m\pi/n$ are sharper when $L$ is divisible by $n$.
Our numerical results definitively suggested an explicit formula for the peak locations and heights, which predicts dense Bragg peaks, as occurs in quasicrystals (e.g., Fibonacci chain \cite{levine1984quasicrystals} as well as other one-dimensional examples \cite{ouguz2017hyperuniformity}), 
but unlike the latter, the peak
locations occur at rational multiples
of $\pi$. 
In an upcoming paper \cite{To17}, we will show that the primes in the intervals studied here are similar to ``limit periodic'' systems \cite{ba11} but still distinct from them. Limit-periodic systems are unions of an infinite number of periodic systems with rational periods, and possess dense Bragg peaks with rational ratios between their locations.
Nevertheless, primes are certainly not unions of periodic systems, since they possess a density gradient, and large primes are difficult to predict \cite{Mersenne}.


In \cite{To17}, we will use the well-known Dirichlet's theorem on arithmetic progressions to show that the primes in the intervals studied here are limit-periodic in a probabilistic sense. 
Based on this theorem, we will provide a number-theoretic explanation for the numerical observations here; specifically, the peak location and height formula, the decrease of peak widths, and how the diffuse part vanishes as $L$ increases.
Moreover, we will show there that if the interval of the primes is chosen to be appreciably smaller or larger than the one used here, the primes would not be characterized by a large degree of order, as measured by the $\tau$ order metric \cite{torquato2015ensemble}.

The numerical techniques that we employed here to investigate
the structure factor of the primes
in finite intervals should be applicable to characterize other complex point configurations
in which the shape and heights
of peaks depend sensitively on the
system size. These examples
include quasicrystals \cite{levine1984quasicrystals} and 
limit-periodic systems \cite{ba11}.
For example, to study the shape and height of a peak located at $k=2\pi/\alpha$, the best choice for the system size is a multiple of $\alpha$. Moreover, the structure factor of any finite system 
with dominant peak structure will
always numerically possess
a ``noisy" diffuse contribution. 
A simple way to characterize the diffuse part is to calculate the median value of $S(k)$ for all allowed $k$ values and then study its
behavior as a function of the system
size to see if it becomes negligible
in the infinite-size limit.

\section*{Acknowledgments}
We are very grateful to Matthew de Courcy-Ireland for helpful discussions.
This work was supported in part by the National Science Foundation under Award No. DMR-1714722.

\section*{References}

\end{document}